\documentclass[12pt, draftclsnofoot, journal, letterpaper, onecolumn]{IEEEtran}

\usepackage[dvips]{graphicx}
\usepackage{cite}
\usepackage{amsmath}
\usepackage{array}
\usepackage{amssymb}

\usepackage{stfloats}
\usepackage{float}
\usepackage[usenames]{color}
\usepackage{xspace,colortbl}

\floatstyle{ruled}
\newfloat{algorithm}{tbp}{loa}
\floatname{algorithm}{Algorithm}

\makeatletter

\begin{document}

\title{A New SLNR-based Linear Precoding for Downlink Multi-User Multi-Stream
MIMO Systems}

\author{Peng Cheng, Meixia Tao and Wenjun Zhang %
\thanks{The authors are with the Department of Electronic Engineering, Shanghai
Jiao Tong University, Shanghai 200240, P. R. China. Emails: \{cp2001cp,
mxtao, zhangwenjun\}@sjtu.edu.cn %
} }

\maketitle
\begin{abstract}
Signal-to-leakage-and-noise ratio (SLNR) is a promising criterion
for linear precoder design in multi-user (MU) multiple-input
multiple-output (MIMO) systems. It decouples the precoder design
problem and makes closed-form solution available. In this letter, we
present a new linear precoding scheme by slightly relaxing the SLNR
maximization for MU-MIMO systems with multiple data streams per
user. The precoding matrices are obtained by a general form of
simultaneous diagonalization of two Hermitian matrices. The new
scheme reduces the gap between the per-stream effective channel
gains, an inherent limitation in the original SLNR precoding scheme.
Simulation results demonstrate that the proposed precoding achieves
considerable gains in error performance over the original one for
multi-stream transmission while maintaining almost the same
achievable sum-rate.
\end{abstract}

\section{Introduction}
\setlength{\arraycolsep}{2pt}

\setlength{\abovedisplayshortskip}{0.15cm}
\setlength{\belowdisplayshortskip}{0.15cm}
\setlength{\abovedisplayskip}{0.15cm}
\setlength{\belowdisplayskip}{0.15cm}

\PARstart{T}{he} significance of a downlink multi-user
multiple-input-multiple-output (MU-MIMO) system is to allow a base
station (BS) to communicate with several co-channel mobile stations
(MS) simultaneously and thereby considerably increase the system
throughput. To utilize the benefit, it is essential to suppress
co-channel interference (CCI). Among many CCI suppression schemes,
linear precoding gains the popularity because of its simplicity for
implementation and good performance. To design the optimal linear
MU-MIMO precoding scheme, it is often desirable to maximize the
output signal-to-interference-plus-noise ratio (SINR) for each user.
However, this problem is known to be challenging due to its coupled
nature and no closed-form solution is available yet. A more
tractable but suboptimal design is to enforce a zero-CCI requirement
for each user, such as block diagonalization (BD) \cite{BD} and
coordinated beamforming (CB) \cite{CB}.

In \cite{SLNR}, the authors propose a so-called
signal-to-leakage-and-noise ratio (SLNR) as the optimization metric
for linear precoder design. This metric transforms a coupled
optimization problem into a completely decoupled one, for which a
closed-form solution is available. Unlike the BD approach, it does
not impose a restriction on the number of transmit antennas at the
BS. Moreover, it is applicable for any number of users and data
streams in contrast to CB scheme. Specifically, the SLNR based
linear precoding weights in \cite{SLNR} are obtained by the
generalized eigenvalue decomposition (GED) of the channel covariance
matrix and the leakage channel-plus-noise covariance matrix of each
user. However, a drawback of such GED based precoding scheme is
that, when each user has multiple data streams, the effective
channel gain for each stream can be severely unbalanced. If power
control or adaptive modulation and coding cannot be applied, the
overall error performance of each user will suffer significant loss.

In this letter, we present a new linear precoding scheme based on
the SLNR criterion for a downlink MU-MIMO system with multiple data
streams per user. The design goal is to reduce the margin between
the effective SINRs of multiple data streams. To do this, we
introduce a slight relaxation for pursuing SLNR maximization (Note
that maximizing SLNR at the transmitter side does not necessarily
lead to output SINR maximization at each receiver). Thereby, we
obtain a general form of simultaneous diagonalization of two
covariance matrices linked to the user\textquoteright{}s channel and
leakage-plus-noise. Based on that, the new precoding matrices are
then obtained. We also present a simple and low-complexity algorithm
to compute the precoding matrix for each user. Simulation results
confirm that, compared with the original scheme, our scheme
demonstrates sizable performance gains in error rate performance for
multi-stream transmission while maintaining almost the same sum-rate
performance.

\emph{Notations}: $\mathrm{\mathbb{E}\left(\cdot\right)}$, $\mathrm{Tr}\left(\cdot\right)$,
$\left(\cdot\right)^{-1}$, and $\left(\cdot\right)^{H}$ denote expectation,
trace, inverse, and conjugate transpose, respectively. $\left\Vert \cdot\right\Vert _{F}$
represents the Frobenius norm. $\mathbf{I}_{N}$ is the $N\times N$
identity matrix. $\mathrm{diag}\left(a_{1},\cdots,a_{N}\right)$ is
the diagonal matrix with element $a_{n}$ on the $n$-th diagonal.
Besides, $\mathbb{C}^{M\times N}$ represents the set of $M\times N$
matrices in complex field.

\section{System Model}

We consider a downlink MU-MIMO system with $N$ transmit antennas and
$M$ receive antennas at each of the $K$ active users. Let
$\mathbf{H}_{k}\in\mathbb{C}^{M\times N}$ denotes the channel from
the BS to the MS $k$ and
$\mathbf{\bar{H}}_{k}=\left[\mathbf{H}_{1}^{H},\cdots,\mathbf{H}_{k-1}^{H},\mathbf{H}_{k+1}^{H},\cdots,\mathbf{H}_{K}^{H}\right]^{H}\in\mathbb{C}^{\left(K-1\right)M\times
N}$ represent the corresponding concatenated leakage channel. A
spatially uncorrelated flat Rayleigh fading channel is assumed. The
elements of $\mathbf{H}_{k}$ are modeled as independent and
identically distributed complex Gaussian variables with zero-mean
and unit-variance. In addition, we assume $\mathbf{H}_{k}$, and also
$\mathbf{\bar{H}}_{k}$, have full rank with probability one. For a
specific vector time, the transmitted vector symbol of user $k$ is
denoted as $\mathbf{s}_{k}\in\mathbb{C}^{L\times1}$, where $L\,(\leq
M)\,$ is the number of data streams supported for user $k$ and is
assumed equal for all the users for simplicity. The vector symbol
satisfies the power constraint
$\mathrm{\mathbb{E}}\left(\mathbf{s}_{k}\mathbf{s}_{k}^{H}\right)=\mathbf{I}_{L}$.
Before entering into the MIMO channel, the vector $\mathbf{s}_{k}$
is pre-multiplied by a precoding matrix
$\mathbf{F}_{k}\in\mathbb{C}^{N\times L}$.
Here, power allocation and rate adaptation among data streams can be
applied. However, the signal design or feedback support may be
relatively complex and thus we resort to precoding design only in
this work. Then, for a given user $k$, the received signal vector
can be written as
 \begin{equation}
\mathbf{r}_{k}=\mathbf{H}_{k}\mathbf{F}_{k}\mathbf{s}_{k}+\mathbf{H}_{k}\sum\nolimits _{i=1,i\ne k}^{K}\mathbf{F}_{i}\mathbf{s}_{i}+\mathbf{n}_{k}
\end{equation}
 in which the second term represents CCI and the third term is the
additive white Gaussian noise with $\mathrm{\mathbb{E}}\left(\mathbf{n}_{k}\mathbf{n}_{k}^{H}\right)=\sigma^{2}\mathbf{I}_{M}$.

We review the original SLNR based precoding scheme in \cite{SLNR}.
Recall that the SLNR is defined as the ratio of received signal power
at the desired MS to received signal power at the other terminals
(the leakage) plus noise power without considering receive matrices,
given by
\begin{equation}
\textnormal{SLNR}_{k}= \frac{\mathrm{Tr}\left(\mathbf{F}_{k}^{H}\mathbf{H}_{k}^{H}\mathbf{H}_{k}\mathbf{F}_{k}\right)}{\mathrm{Tr}\left(\mathbf{F}_{k}^{H}\left(M/L\sigma^{2}\mathbf{I}+\mathbf{\bar{H}}_{k}^{H}\mathbf{\bar{H}}_{k}\right)\mathbf{F}_{k}\right)},
\end{equation}
 for $k=1,\cdots,K.$ According to the SLNR criterion, the precoding
matrix $\mathbf{F}_{k}$ is designed based on the following metric
\begin{equation}
\mathbf{F}_{k}^{\rm{opt}}=\arg\underset{\mathbf{F}_{k}\in\mathbb{C}^{N\times
L}}{\mathop{\max}}\,\textnormal{SLNR}_{k}\label{eq:SLNRmax}
\end{equation}
 with $\textnormal{Tr}\left(\mathbf{F}_{k}\mathbf{F}_{k}^{H}\right)=L$
for power limitation. Since $\mathbf{H}_{k}^{H}\mathbf{H}_{k}$ is
Hermitian and positive semidefinite (HPSD) and $M/L\sigma^{2}\mathbf{I}+\mathbf{\bar{H}}_{k}^{H}\mathbf{\bar{H}}_{k}$
is Hermitian and positive definite (HPD), by generalized eigenvalue
decomposition, there exists an invertible matrix $\mathbf{T}_{k}\in\mathbb{C}^{N\times N}$
such that
\begin{eqnarray}
\mathbf{T}_{k}^{H}\mathbf{H}_{k}^{H}\mathbf{H}_{k}\mathbf{T}_{k}=\mathbf{\mathbf{\Lambda}}_{k} & = & \textnormal{diag}\left(\lambda_{1},\cdots,\lambda_{N}\right)\label{eq:diag1_sub}\\
\mathbf{T}_{k}^{H}\left(M/L\sigma^{2}\mathbf{I}+\mathbf{\bar{H}}_{k}^{H}\mathbf{\bar{H}}_{k}\right)\mathbf{T}_{k} & = & \mathbf{I}_{N}\label{eq:diag1}
\end{eqnarray}
 with $\lambda_{1}\ge\lambda_{2}\ge\cdots\ge\lambda_{N}\ge0$. Here,
the columns of $\mathbf{T}_{k}$ and the diagonal entries of
$\mathbf{\mathbf{\Lambda}}_{k}$ are the generalized eigenvectors and
eigenvalues of the pair $\left\{ \mathbf{H}_{k}^{H}\mathbf{H}_{k},\:
M/L\sigma^{2}\mathbf{I}+\mathbf{\bar{H}}_{k}^{H}\mathbf{\bar{H}}_{k}\right\}
$, respectively. It is then shown in \cite{SLNR} that the optimal
precoder which is able to maximize the objective function
(\ref{eq:SLNRmax}) can be obtained by extracting the leading $L$
columns of $\mathbf{T}_{k}$ as
\begin{equation}
\mathbf{F}_{k}^{\rm{opt}}=\rho\mathbf{T}_{k}\left[\mathbf{I}_{L};\mathbf{0}\right],\label{eq:F_k}
\end{equation}
where $\rho$ is a scaling factor so that
$\textnormal{Tr}\left(\mathbf{F}_{k}\mathbf{F}_{k}^{H}\right)=L$.
The resulting maximum SLNR value is given by
$\textnormal{SLNR}_{k}^{\mathrm{max}}=\sum_{i=1}^{L}\lambda_{i}/L$.
Along with the realization of the precoder, the matched-filter type
receive matrix, denoted as
$\mathbf{G}_{k}=\left(\mathbf{H}_{k}\mathbf{F}_{k}\right)^{H}$, is
applied at each user receiver, resulting in inter-stream
interference free.
Note that better performance could be achieved if a multi-user MMSE
type receiver is adopted. In this letter, we still adopt MF-type
detector at the receiver as in [3] for implementation simplicity and
analytical convenience.
%
%

A drawback of such GED based precoding scheme is that, when $L\ge2$,
the effective channel gain for each stream can be severely
unbalanced as shall be illustrated in Section III-C. It is known
that the overall performance of a user with multiple streams is
dominated by the stream with the worst channel condition. Hence,
such channel imbalance would lead to poor overall error performance
for a user. In the next section, we allow a slight relaxation on the
SLNR maximization, which provides additional degrees of freedom to
design a new precoding scheme so as to overcome this drawback.

\section{Proposed Precoding Scheme}

\subsection{Design Principle by Matrix Theory}

The expressions in (\ref{eq:diag1_sub}) and (\ref{eq:diag1}) by
the GED approach motivate us to find a more general form of simultaneous
diagonalization of two matrices. Before introducing our results in
Proposition 1, we review the following Lemma~\cite[Ch. 4, 4.5.8]{Horn}:

\textbf{Lemma 1}: Let $\mathbf{A},\,\mathbf{B}\in\mathbb{C}^{n\times n}$
be Hermitian. There is a non-singular matrix $\mathbf{S}\in\mathbb{C}^{n\times n}$
such that $\mathbf{S}^{H}\mathbf{A}\mathbf{S}=\mathbf{B}$ if and
only if $\mathbf{A}$ and $\mathbf{B}$ have the same inertia, that
is, have the same number of positive, negative, and zero eigenvalues.

\textbf{Proposition 1}: For the pair of matrices $\left\{ \mathbf{H}_{k}^{H}\mathbf{H}_{k},\: M/L\sigma^{2}\mathbf{I}+\mathbf{\bar{H}}_{k}^{H}\mathbf{\bar{H}}_{k}\right\} $,
there is a non-singular matrix $\mathbf{P}_{k}\in\mathbb{C}^{N\times N}$
such that
\begin{eqnarray}
\mathbf{P}_{k}^{H}\mathbf{H}_{k}^{H}\mathbf{H}_{k}\mathbf{P}_{k} & = & \mathbf{\Theta}_{k}\label{eq:diag2_sub}\\
\mathbf{P}_{k}^{H}\left(M/L\sigma^{2}\mathbf{I}+\mathbf{\bar{H}}_{k}^{H}\mathbf{\bar{H}}_{k}\right)\mathbf{P}_{k} & = & \mathbf{\Omega}_{k}\label{eq:diag2}
\end{eqnarray}
 in which $\mathbf{\Theta}_{k}=\textnormal{diag}\left(\theta_{1},\theta_{2},\cdots\theta_{N}\right)$
and $\mathbf{\Omega}_{k}=\textnormal{diag}\left(\omega_{1},\omega_{2},\cdots\omega_{N}\right)$
with the entries satisfying $1>\theta_{1}\ge\cdots\ge\theta_{M}>0,\,\,\theta_{M+1}=\cdots=\theta_{N}=0$
and $0<\omega_{1}\le\cdots\le\omega_{M}<1,\,\,\omega_{M+1}=\cdots=\omega_{N}=1$
as well as $\theta_{i}+\omega_{i}=1$ for $i=1,2,\cdots,N$.

\begin{proof}
Denote $\mathbf{A}_{k}=\mathbf{H}_{k}^{H}\mathbf{H}_{k}$, $\mathbf{B}_{k}=M/L\sigma^{2}\mathbf{I}+\mathbf{\bar{H}}_{k}^{H}\mathbf{\bar{H}}_{k}$
and $\mathbf{C}_{k}=\mathbf{A}_{k}+\mathbf{B}_{k}$. Let the eigenvalues
$\lambda_{i}\left(\mathbf{A}_{k}\right),$ $\lambda_{i}\left(\mathbf{B}_{k}\right)$
and $\lambda_{i}\left(\mathbf{C}_{k}\right)$, $i=1,2,\cdots,N$,
be arranged in increasing order. Since $\mathbf{A}_{k}$ is HPSD and
$\mathbf{B}_{k}$ is HPD, namely, $\lambda_{i}\left(\mathbf{A}_{k}\right)\geq0$
and $\lambda_{i}\left(\mathbf{B}_{k}\right)>0$, then by \cite[4.3.1]{Horn},
we have $\lambda_{i}\left(\mathbf{C}_{k}\right)\ge\lambda_{i}\left(\mathbf{A}_{k}\right)+\lambda_{1}\left(\mathbf{B}_{k}\right)>0$,
$\forall i$. This implies that $\mathbf{C}_{k}$ is HPD. Then, by
the matrix theory in \cite[4.5.8, Exercise]{Horn}, there must be
a non-singular matrix $\mathbf{Q}_{k}\in\mathbb{C}^{N\times N}$ such
that
\begin{equation}
\mathbf{Q}_{k}^{H}\mathbf{C}_{k}\mathbf{Q}_{k}=\mathbf{Q}_{k}^{H}\left(\mathbf{A}_{k}+\mathbf{B}_{k}\right)\mathbf{Q}_{k}\mathbf{=I}_{N}.\label{eq:In}
\end{equation}
 Further, denote $\mathbf{A}_{k}^{\prime}=\mathbf{Q}_{k}^{H}\mathbf{A}_{k}\mathbf{Q}_{k}$
and $\mathbf{B}_{k}^{\prime}=\mathbf{Q}_{k}^{H}\mathbf{B}_{k}\mathbf{Q}_{k}$.
By Lemma 1, it can be shown that $\mathbf{A}_{k}^{\prime}$ and $\mathbf{B}_{k}^{\prime}$
have the same inertia with $\mathbf{A}_{k}$ and $\mathbf{B}_{k}$,
respectively. Thus, $\mathbf{A}_{k}^{\prime}$ is HPSD and $\mathbf{B}_{k}^{\prime}$
is HPD. Now, by using \cite[4.3.1]{Horn} again, it is easy to show
that $1>\lambda_{i}\left(\mathbf{A}_{k}^{\prime}\right)\geq0$ and
$1\geq\lambda_{i}\left(\mathbf{B}_{k}^{\prime}\right)>0$. Next, according
to the eigen-decomposition (ED) of a Hermitian matrix \cite{Horn},
there must be a unitary matrix $\mathbf{U}_{k}\in\mathbb{C}^{N\times N}$
such that
\begin{equation}
\mathbf{U}_{k}^{H}\mathbf{A}_{k}^{\prime}\mathbf{U}_{k}=\mathrm{diag}\left(\lambda_{1}\left(\mathbf{A}_{k}^{\prime}\right),\cdots,\lambda_{N}\left(\mathbf{A}_{k}^{\prime}\right)\right).\label{eq:U}
\end{equation}
 Applying $\mathbf{U}_{k}$ in both sides of (\ref{eq:In}), we obtain
\begin{equation}
\mathbf{U}_{k}^{H}\left(\mathbf{A}_{k}^{\prime}+\mathbf{B}_{k}^{\prime}\right)\mathbf{U}_{k}\mathbf{=I}_{N}.\label{In2}\end{equation}
 Hence, observing (\ref{eq:U}) and (\ref{In2}), we find that it
is necessary for
$\mathbf{U}_{k}^{H}\mathbf{B}_{k}^{\prime}\mathbf{U}_{k}$ to satisfy
$\mathbf{U}_{k}^{H}\mathbf{B}_{k}^{\prime}\mathbf{U}_{k}=\mathrm{diag}\left(\left(1-\lambda_{1}(\mathbf{A}_{k}^{\prime})\right),\cdots,\left(1-\lambda_{N}(\mathbf{A}_{k}^{\prime})\right)\right).$
Clearly, as $\mathbf{U}_{k}$  is unitary, then $\left\{
1-\lambda_{i}\left(\mathbf{A}_{k}^{\prime}\right)\right\}
_{i=1}^{N}$ must be the eigenvalues of $\mathbf{B}_{k}^{\prime}$. To
this end, we define $\mathbf{P}_{k}=\mathbf{Q}_{k}\mathbf{U}_{k}$.
Since
$\mathrm{rank}\left(\mathbf{H}_{k}^{H}\mathbf{H}_{k}\right)=\mathrm{rank}\left(\mathbf{H}_{k}\right)=M$
and the rank is unchanged upon left or right multiplication by a
nonsingular matrix, then we arrive at the results in
(\ref{eq:diag2_sub}) and (\ref{eq:diag2}).
\end{proof}
\begin{algorithm}
{\footnotesize \caption{{\small The specific design of precoder $\mathbf{F}_{k}^{\mathbf{\prime}}$
for user $k$ }}
}{\footnotesize \par}

{\footnotesize Input: $\mathbf{A}_{k}=\mathbf{H}_{k}^{H}\mathbf{H}_{k}$,
and $\mathbf{C}_{k}=\left(\mathbf{H}_{k}^{H}\mathbf{H}_{k}+M/L\sigma^{2}\mathbf{I}+\mathbf{\bar{H}}_{k}^{H}\mathbf{\bar{H}}_{k}\right)$ }{\footnotesize \par}

{\footnotesize 1) Compute Cholesky decomposition on $\mathbf{\mathbf{C}}_{k}$,
as $\mathbf{\mathbf{C}}_{k}=\mathbf{G}_{k}\mathbf{G}_{k}^{H},$ where
$\mathbf{\mathbf{G}}_{k}\in\mathbb{C}^{N\times N}$ is a lower triangular
matrix with positive diagonal entries. Then, $\mathbf{\mathbf{G}}_{k}^{-1}$
can be easily obtained and we have $\left(\mathbf{\mathbf{G}}_{k}^{-1}\right)^{H}=\mathbf{Q}_{k}$
in (\ref{eq:In}). }\\
{\footnotesize 2) Compute $\mathbf{A}_{k}^{\prime}=\mathbf{Q}_{k}^{H}\mathbf{A}_{k}\mathbf{Q}_{k}$,
then compute ED on $\mathbf{A}_{k}^{\prime}$ as $\mathbf{A}_{k}^{\prime}\mathbf{U}_{k}=\mathbf{U}_{k}\mathbf{\Lambda}_{k}$.
Note $\mathbf{U}_{k}$ must be unitary and it can be also obtained
by computing the left singular matrix of $\mathbf{A}_{k}^{\prime}$
in terms of SVD. }\\
{\footnotesize 3) Compute $\mathbf{P}_{k}=\mathbf{Q}_{k}\mathbf{U}_{k}$.
}\\
{\footnotesize Output: }{\small $\mathbf{F}_{k}^{\mathbf{\prime}}=\gamma\mathbf{P}_{k}\left(\mathbf{I}_{L};\mathbf{0}\right)$.}
\end{algorithm}

\subsection{Precoder Design}

The simultaneous diagonalization in general form stated in
Proposition 1 draws a significant distinction from the original GED
based deduction in (\ref{eq:diag1_sub}) and (\ref{eq:diag1}). This
allows us to design a new precoding scheme. In specific, the
proposed precoder $\mathbf{F}_{k}^{\mathbf{\prime}}$ and matched
decoder $\mathbf{G_{\mathnormal{k}}^{\mathbf{\prime}}}$ can be
designed as
\begin{equation}
\mathbf{F}_{k}^{\mathbf{\prime}}=\gamma\mathbf{P}_{k}\left[\mathbf{I}_{L};\mathbf{0}\right],\quad\mathbf{G_{\mathnormal{k}}^{\mathbf{\prime}}}=\left(\mathbf{H}_{k}\mathbf{F}_{k}^{\mathbf{\prime}}\right)^{H}\label{eq:F_k'}
\end{equation}
 in which $\gamma$ is a normalization factor so that $\textnormal{Tr}\left(\mathbf{F}_{k}^{\mathbf{\prime}}\mathbf{F}_{k}^{\mathbf{\prime}H}\right)=L$.
It is clear that $\mathbf{G_{\mathnormal{k}}^{\mathbf{\prime}}}\mathbf{H}_{k}\mathbf{F}_{k}$
amounts to a certain diagonal matrix, also resulting in inter-stream-interference
free.

The remaining problem is how to compute a specific precoder
$\mathbf{F}_{k}^{\mathbf{\prime}}$ for each user. Based on our proof
of Proposition 1, we present a closed-form expression using a simple
and low-complex algorithm, outlined in Algorithm 1. In the next
subsection, we reveal the superiority of the proposed precoding
scheme through per-stream SINR discussion.


\subsection{Performance Discussion}


Firstly, continuing to use the same symbols $\mathbf{A}_{k}$ and
$\mathbf{B}_{k}$ as in the proof of Proposition 1, we can show that
$\mathbf{A}_{k}\mathbf{t}_{k_{i}}=\lambda_{i}\mathbf{B}_{k}\mathbf{t}_{k_{i}}$
and $\mathbf{A}_{k}\mathbf{p}_{k_{i}}=\left(\theta_{i}/\omega_{i}\right)\mathbf{B}_{k}\mathbf{p}_{k_{i}}$
from (\ref{eq:diag1}) and (\ref{eq:diag2}), in which $\mathbf{t}_{k_{i}}$
and $\mathbf{p}_{k_{i}}$ correspond to the $i$-th column of $\mathbf{T}_{k}$
and $\mathbf{P}_{k}$, respectively. Here, both $\lambda_{i}$ and
$\theta_{i}/\omega_{i}$ must be the generalized eigenvalues of the
pair $\left\{ \mathbf{A}_{k},\mathbf{B}_{k}\right\} $. It is then
easy to see that
\begin{equation}
\lambda_{j}=\theta_{j}/\omega_{j},\; j=1,2,\cdots,N\label{eq:lamda}
\end{equation}
 with $\left\{ \lambda_{j}\right\} _{j=1}^{N}$ and $\left\{ \theta_{j}\right\} _{j=1}^{N}$
being sorted in descending order while $\left\{ \omega_{j}\right\}
_{j=1}^{N}$ sorted in ascending order. Now we have
$\textnormal{SLNR}_{k}=\left(\sum_{l=1}^{L}\theta_{l}\right)/\left(\sum_{l=1}^{L}(1-\theta_{l})\right)$,
which is slightly smaller than
$\textnormal{SLNR}_{k}^{\mathrm{max}}$ given in Section II.

On the other hand, the ultimate performance is decided by post-SINR.
Clearly, the decoded signal should take the form
\begin{equation}
\mathbf{\hat{s}}_{k}=\mathbf{G}_{k}^{\prime}\mathbf{H}_{k}\mathbf{F}_{k}\mathbf{s}_{k}+\mathbf{G}_{k}^{\prime}\left(\mathbf{H}_{k}\sum\nolimits _{i=1,i\ne k}^{K}\mathbf{F}_{i}\mathbf{s}_{i}+\mathbf{n}_{k}\right).
\end{equation}
 Thanks to diagonal form in (\ref{eq:diag2_sub}) and (\ref{eq:diag2}), the covariance matrix
of noise vector is given by
$\mathrm{\mathbb{E}}\left(\mathbf{G}_{k}^{\prime}\mathbf{n}_{k}\mathbf{n}_{k}^{H}\mathbf{G}_{k}^{\prime
H}\right)=\sigma^{2}\mathbf{I}_{L}\mathbb{E}\left(\mathbf{G}_{k}^{\prime}\mathbf{G}_{k}^{\prime
H}\right)=\textrm{\ensuremath{\gamma^{2}\sigma^{2}\mathbf{I}_{\mathit{L}}\mathrm{diag}\left(\theta_{1},\cdots,\theta_{L}\right).}}$
Furthermore, it can be verified through numerical results (difficult
via theoretical analysis though) that the residual CCI is much
smaller than the noise power at high SNR. As such, the SINR on the
$l$-th stream, $\eta_{l}^{\prime}$ can be approximately calculated
as
$\eta_{l}^{\prime}=\left(\gamma^{4}\theta_{l}^{2}\right)/\left(\gamma^{2}\sigma^{2}\theta_{l}\right)=\gamma^{2}\theta_{l}/\sigma^{2}$
. Then, for any two streams $l$ and $m$ with $l>m$, the margin of
$\Delta_{l,m}^{\prime}$ between $\eta_{l}^{\prime}$ and
$\eta_{m}^{\prime}$ in terms of decibel (dB) can be expressed as
\begin{equation}
\Delta_{l,m}^{\prime}=10\mathrm{log}_{10}\left(\eta_{l}^{\prime}/\eta_{m}^{\prime}\right)=10\mathrm{log}_{10}\left(\theta_{l}/\theta_{m}\right).
\end{equation}
 Following the same analysis, the margin of $\Delta_{l,m}$ for the
original scheme can be analogously calculated as
\begin{equation}
\Delta_{l,m}=10\mathrm{log}_{10}\left(\eta_{l}/\eta_{m}\right)=10\mathrm{log}_{10}\left(\lambda_{l}/\lambda_{m}\right).
\end{equation}
 According to (\ref{eq:lamda}), we have $\lambda_{l}/\lambda_{m}=\left(\theta_{l}\omega_{m}\right)/\left(\theta_{m}\omega_{l}\right)$.
Further, we have that $\omega_{m}>\omega_{l}$ for $l>m$ by definition.
It then ensures that the following inequality holds:
\begin{equation}
\Delta_{l,m}^{\prime}<\Delta_{l,m}.\label{eq:deta decrease}
\end{equation}
 This explicitly shows that the SINR margin between any two streams
decreases by applying the proposed scheme. In other words, the effective
channel gains between the multiple streams are now less unbalanced.
Its effectiveness will be further examined by simulation in the next
section.



%
\begin{figure}
\begin{centering}
\vspace{-1cm}

\par\end{centering}

\begin{centering}
\includegraphics[scale=0.8]{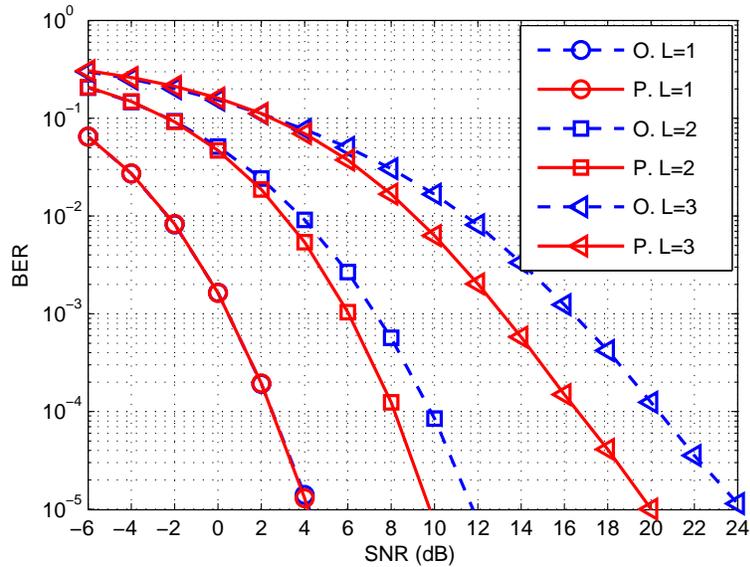} \vspace{-0.3cm}
 \caption{Uncoded BER of a MU-MIMO system with $N=8$ transmit
antennas at the BS and $K=2$ users each with $M=3$ receive
antennas.}

\par\end{centering}

\begin{centering}
\label{fig:BER}
\par\end{centering}

\vspace{-0.4cm}

\end{figure}


\section{Simulation Results}

Fig.~1 compares the simulated bit error rate (BER) per user in a
MU-MIMO system with different system configurations. Here, P denotes
the proposed precoding scheme and O denotes the original scheme in
\cite{SLNR}. QPSK modulation with Gray mapping is employed and the
BER curves are plotted versus the transmit SNR ($L/\sigma^{2})$. It
is seen that the proposed scheme and the original scheme for
single-stream case ($L=1$) achieve the same BER performance. For
multiple streams ($L=2$ and 3), the former outperforms the latter
with sizeable gains. In specific, a gain of around $2$ dB and $4$ dB
can be achieved at BER=$10^{-4}$ for streams of $L=2$ and $L=3$,
respectively. We also carried out the achievable sum-rate
comparison. It is found that our scheme is almost the same as the
original one. The results are omitted due to page limit.

The above simulation results verify the effectiveness of the
proposed precoding scheme over the original SLNR based scheme when
there are multiple data streams for each user.



\end{document}